\documentclass[letter]{aa}  

\usepackage{graphicx}
\usepackage{caption}
\usepackage{subcaption}
\graphicspath{{./figures/}}
\usepackage{txfonts}
\usepackage{hyperref}
\hypersetup{
  colorlinks   = true, 
  urlcolor     = blue, 
  linkcolor    = blue, 
  citecolor   = blue 
}
\begin{document}

   \title{Four new eclipsing accreting ultracompact white dwarf binaries found with the Zwicky Transient Facility}


   \author{J.~M. Khalil \inst{1}
          \and J. van Roestel\inst{1}\fnmsep\thanks{j.c.j.vanroestel@uva.nl}
          \and\\ E.~C. Bellm\inst{2} 
          \and J.~S. Bloom\inst{3,4}  
          \and R. Dekany \inst{5}
          \and A.~J. Drake\inst{6}
          \and M.~J. Graham\inst{6}
          \and S.~L. Groom\inst{7}
          \and S.~R. Kulkarni\inst{6}
          \and R.~R. Laher\inst{7}
          \and A.~A. Mahabal\inst{6}
          \and T. Prince\inst{6}
          \and R. Riddle\inst{5}
          }

   \institute{Anton Pannekoek Institute for Astronomy, University of Amsterdam, 1090 GE Amsterdam, The Netherlands
   \and 
   DIRAC Institute, Department of Astronomy, University of Washington, 3910 15th Avenue NE, Seattle, WA 98195, USA;    
   \and
    Department of Astrophysics, University of California, Berkeley, CA 94720-3411, USA
    \and
    Lawrence Berkeley National Laboratory, 1 Cyclotron Road, MS 50B-4206, Berkeley, CA 94720, USA
    \and
    Caltech Optical Observatories, California Institute of Technology, Pasadena, CA  91125, USA
    \and 
    Division of Physics, Mathematics, and Astronomy, California Institute of Technology, Pasadena, CA 91125, USA
   \and
    IPAC, California Institute of Technology, 1200 E. California Boulevard, Pasadena, CA 91125, USA
    }
   \date{Received December 15, 2023; accepted ??? ??, ????}

 
  \abstract
   {Accreting ultracompact binaries contain a white dwarf that is accreting from a degenerate object and have orbital periods shorter than 65 minutes.}
{The aims of this letter are to report the discovery and the orbital period of four new eclipsing accreting ultracompact binaries found using the Zwicky Transient Facility, and to discuss their photometric properties.}
{We searched through a list of 4171 dwarf novae compiled using the Zwicky Transient Facility and used the Box Least Square method to search for periodic signals in the data.}
{We found four new eclipsing accreting ultracompact binaries with orbital periods between 25.9--56 minutes, one of which is previously published as an AM CVn, while the other three systems are new discoveries. The other two shorter period systems are likely also AM CVn systems, while the longest period system with a period of 56 minutes shows multiple super-outbursts observed in two years which is more consistent with it being a Helium-CV.}
{}

   \maketitle
%

\section{Introduction}
Accreting ultracompact binaries are white dwarfs that are stably accreting from a degenerate or semi-degenerate donor. The most common subtype are the AM CVn-type systems that do not show any hydrogen and consist mostly of Helium; for recent overview articles, see \citet{2010PASP..122.1133S} and \citet{2018A&A...620A.141R}. Orbital periods of AM CVn stars range from 5 to about 65 minutes. This means that their orbital evolution is governed by gravitational wave radiation and short-period AM CVn systems will be detectable by the LISA satellite \citep[e.g.][]{2023arXiv230212719K}. 

Other types of accreting ultracompact binary systems are helium cataclysmic variables. These are less common compared to AM CVn systems and can also have short orbital periods. What distinguishes them from AM CVn systems is that they still have detectable hydrogen in their atmospheres. 
 Examples are CSS 120422:J111127+571239 with a period of 55.3 minutes \citep{2012ATel.4112....1G}; CRTS J112253.3-111037 with a period of 65.2 minutes \citep{2012MNRAS.425.2548B}; V418 Ser with a 65.9 minute period and CRTS J1028-0819 with a 52.1 minute period \citep{green2020}. In most of these systems the donor is cold, but there are a few exceptions where the donor is warm and can be detected in the optical \citep[e.g.][]{2022Natur.610..467B}.

The formation channel of the different accreting ultracompact binaries is uncertain, which is one of the large open issues in the broader context of compact binary evolution. The formation channels involve a sequence of binary interactions: one or more common-envelope events and/or phases of stable mass transfer. Understanding the evolutionary history of accreting ultracompact binaries is part of the larger question of the final fate of (merging) white dwarf binary stars \citep[see the Horizon-2020 paper ][]{2019BAAS...51c.168T}.

The number of known accreting ultracompact binaries is small but is steadily increasing \citep{green2023} due to large-scale surveys such as the Zwicky Transient Facility \cite[ZTF; ][]{2019PASP..131a8003M,2019PASP..131a8002B,2019PASP..131g8001G,2020PASP..132c8001D}. ZTF is an optical time-domain survey that images the visible sky every 2--3 nights since 2018. It has already been used to detect dwarf novae \citep{2020AJ....159..198S,2021AJ....162...94S}, including dwarf novae associated with accreting ultracompact binaries \citep{2021AJ....162..113V}. Additionally, ZTF lightcurves of stars are also analysed to find short-period variability \citep{2020ApJ...905...32B}, and have been used to find new eclipsing AM CVn stars \citep{2022MNRAS.512.5440V}. In addition to optical photometric surveys, X-ray missions such as ROSAT and SRG/eROSITA have been used to identify AM CVns, including eclipsing systems \citep{2023ApJ...954...63R}.
Eclipsing accreting ultracompact binaries are particularly useful because, with only a high-speed lightcurve, all binary parameters can be measured \citep{2011MNRAS.410.1113C,2018MNRAS.476.1663G,2022MNRAS.512.5440V}.

In this Letter, we present the discovery of four new eclipsing accreting ultracompact binaries that were initially detected as dwarf novae outbursts by ZTF. Section~\ref{sec:data} briefly describes the selection of dwarf novae and the ZTF lightcurves. In Section~\ref{sec:method} we discuss how the ZTF lightcurves were analysed and how the new systems were discovered. We summarise the results in Section~\ref{sec:results} and discuss the results in Section~\ref{sec:Discussion}. We conclude this letter in Section~\ref{sec:conclusion}.

\section{Data}\label{sec:data}
    As a list of targets, we use all dwarf novae identified using ZTF. This list is compiled by human scanners who frequently check the ZTF alert stream for stars that brighten by 1.5 magnitudes using the \textit{Fritz} marshall \citep{vanderWalt2019,coughlin2023}. For a detailed description of the selection of dwarf novae, see \citet{2020AJ....159..198S,2021AJ....162...94S,2021AJ....162..113V,2023AAS...24145406S}. The sample of dwarf novae we analysed contains a total of 4171 objects.

We used the ZTF forced-photometry service to obtain the lightcurve of each object \citep{2023arXiv230516279M}. Low-quality data was removed on the basis of the conditions mentioned by \cite{2023arXiv230516279M}. In addition, we also removed epochs with anomalous zeropoints. The criteria are as follows:
\begin{itemize}
    \item $\mathrm{infobitssci} \geq 33554432$
    \item $\mathrm{scisigpix} > 25$
    \item $\mathrm{sciinpseeing}  > 4''$
    \item $\mathrm{nearestrefsharp}$ is not $null$
    \item $\mathrm{zpdiff} > 26.5$
\end{itemize}
The rejected epochs either have bad calibration issues or are unusable due to high noise.

\section{Method and Analysis}\label{sec:method}
 We searched through the ZTF lightcurve to find eclipses with periods less than 65 minutes using the box-least-squares method (BLS, \citealt{kovacs2002}). The BLS method is an algorithm that searches for periodic eclipse-like signals in time-series data and provides the best estimate for the period. Because accreting ultracompact binaries tend to have very deep eclipses and do not show much other periodic variability, the BLS algorithm is well suited to find the periodic eclipses of accreting ultracompact binaries \citep[see also][]{2022MNRAS.512.5440V}.
 The BLS implementation that we used comes from the \textsc{astrobase} package \citep{wbhatti_astrobase}. We restricted our search to periods of 4.32--288 minutes and used a linear grid in frequency, with a fixed frequency step size of $10^{-5}$ per day. The frequency stepsize is determined by the minimum eclipse width in frequency (0.02) and the time baseline of the lightcurve (5~years).
 We used the built-in sigma-clipping function of the astrobase package to remove outbursts in the data, since they confuse the BLS algorithm. To do this, we used sigma-clipping values of [99, 3].
 
The "HELIOS" computer cluster was used to run our code on the 4171 lightcurves. We visually inspected the output and identified strong peaks in the periodogram at short periods (65 minutes or less). We then looked at the folded lightcurves to identify clear eclipses and found four systems that showed clear eclipses.


\section{Results}\label{sec:results}

\begin{table*}
\tiny 
\caption{Properties of the four new eclipsing AM CVn systems.} 
\label{tab:overview}
\renewcommand{\arraystretch}{1.25}
\begin{tabular}{llllllllll}
Name & RA & Dec & $P_\mathrm{orb}$ & $G$ & $BP-RP$ & $PS$ $g$-$r$ & $PS$ $g$-$z$ & parallax & distance  \\
& & & minutes & Vega-mag & Vega-mag & AB-mag & AB-mag & mas & pc  \\
\hline\hline

ZTF20aabowdt & 07$^{\rm h}$29$^{\rm m}$07.69$^{\rm s}$ & 
\textminus06$^{\circ}$02$^{'}$46.61$^{''}$& 25.92 &20.41 & 
0.07$\pm$0.15 & \phantom{+}0.33$\pm$0.24 &\textminus0.30$\pm$0.09 & 1.17$\pm$0.64 &$2029^{+3612}_{-930}$ \\

ZTF18acgmwpt & 07$^{\rm h}$01$^{\rm m}$15.77$^{\rm s}$ & 
+50$^{\circ}$23$^{'}$21.46$^{''}$ &32.47 &20.50 & 0.14$\pm$0.28 &\textminus0.24$\pm$0.04 &\textminus0.54$\pm$0.05 & 0.50$\pm$0.62& $1636^{+1289}_{-668}$\\

ZTF19abugzba & 18$^{\rm h}$38$^{\rm m}$47.16$^{\rm s}$ & 
+07$^{\circ}$44$^{'}$46.23$^{''}$& 39.61 &20.07 &
0.64$\pm$0.21 & \phantom{+}0.03$\pm$0.06 & \phantom{+}0.10$\pm$0.14 & 1.43$\pm$0.54 & $1740^{+1692}_{-851}$ \\

ZTF21abbxnbm & 21$^{\rm h}$16$^{\rm m}$04.73$^{\rm s}$ & 
+24$^{\circ}$46$^{'}$20.53$^{''}$&56.16 &20.64 & 2.36$\pm$0.88 & \textminus0.09$\pm$0.08 & \phantom{+}0.24$\pm$0.10 &--&-- \\

\hline
\end{tabular}
\tablefoot{The \textit{Gaia} magnitude and parallax are taken from \cite{2023A&A...674A...1G} and \cite{2016A&A...595A...1G}. The Pan-STARRS data was taken from \cite{2016arXiv161205560C}. Distances are taken from \cite{2018AJ....156...58B}.}
\end{table*}

We found four eclipsing accreting ultracompact binaries in our sample of ZTF dwarf novae, see Table \ref{tab:overview}. ZTF20aabowdt has been identified as a white dwarf candidate by \cite{2021MNRAS.508.3877G} using \textit{Gaia} data, and we identified it as an AM CVn system with an orbital period of 25.92 minutes. 
ZTF18acgmwpt has already been published by \cite{2021AJ....162..113V}, they confirmed that it is an AM CV system with He absorption lines in its spectrum. 
Using the super-outburst recurrence time, they estimated that the orbital period should be between 30.7--34.5 minutes. We confirmed that the system period is 32.47 minutes. ZTF19abugzba and ZTF21abbxnbm are new discoveries that have not been reported in the literature; they have orbital periods of 39.61 and 56.16 minutes, respectively. 

All systems are faint with \textit{Gaia} $G$-mag values varying between 20.07 and 20.64. The faintest system (ZTF21abbxnbm) does not have a parallax measurement in \textit{Gaia} and therefore has no information on its distance. 

Accreting ultracompact binaries systems are typically blue because the donor is cold and is not visible in the near-infrared \citep[e.g. Fig. 11 in][]{2022MNRAS.512.5440V}.
ZTF20aabowdt and ZTF18acgmwpt have a blue colour, evident from their \textit{BP--RP} values of 0.07 and 0.14, respectively, and their \mbox{\textit{PS (g-z)}} values of \textminus0.3 and \textminus0.54, respectively. ZTF19abugzba and ZTF21abbxnbm have redder colours with \textit{BP--RP} value of 0.64 and 2.36, respectively. However, the precision of the \textit{BP--RP} colour is low due to the faint nature of the systems, especially ZTF21abbxnbm, rendering some \textit{BP--RP} values unreliable. We therefore compare the \textit{Gaia} colours with the \mbox{\textit{PS (g--z)}} values. These are more precise and do not suggest an anomalously red colour.

\begin{figure*}[h!]
\includegraphics[width = 9 cm]{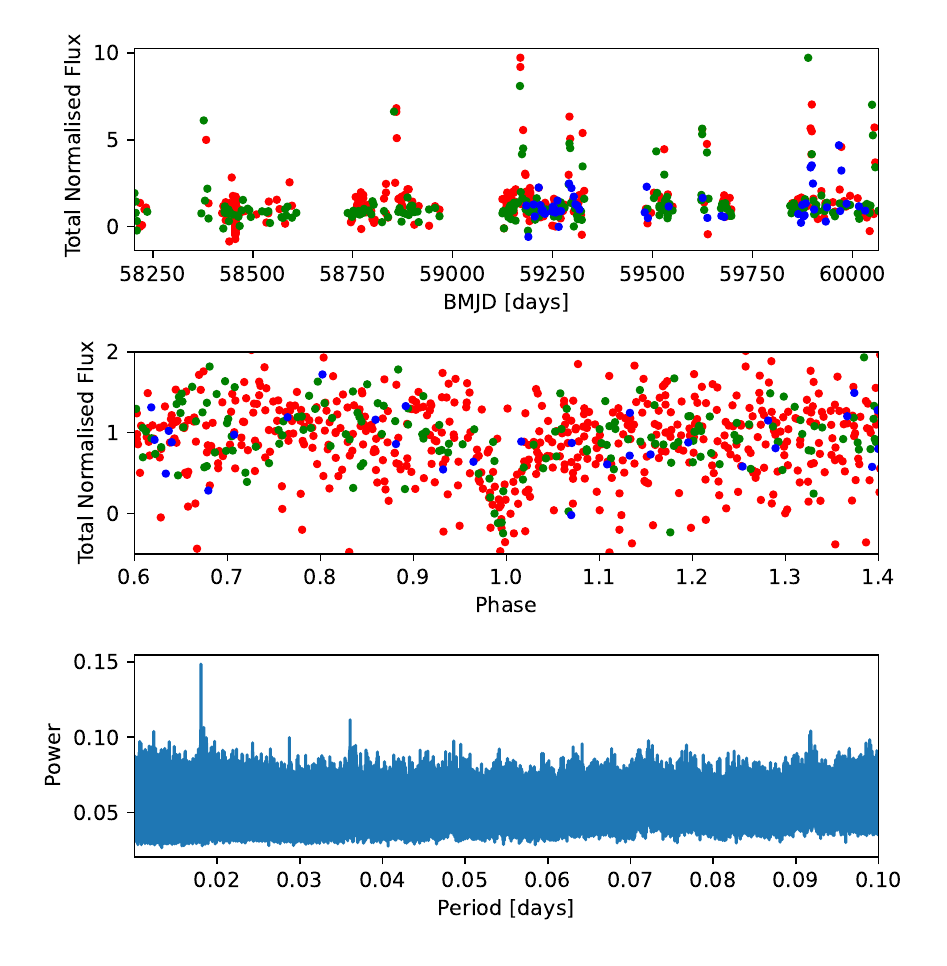}
\includegraphics[width = 9 cm]{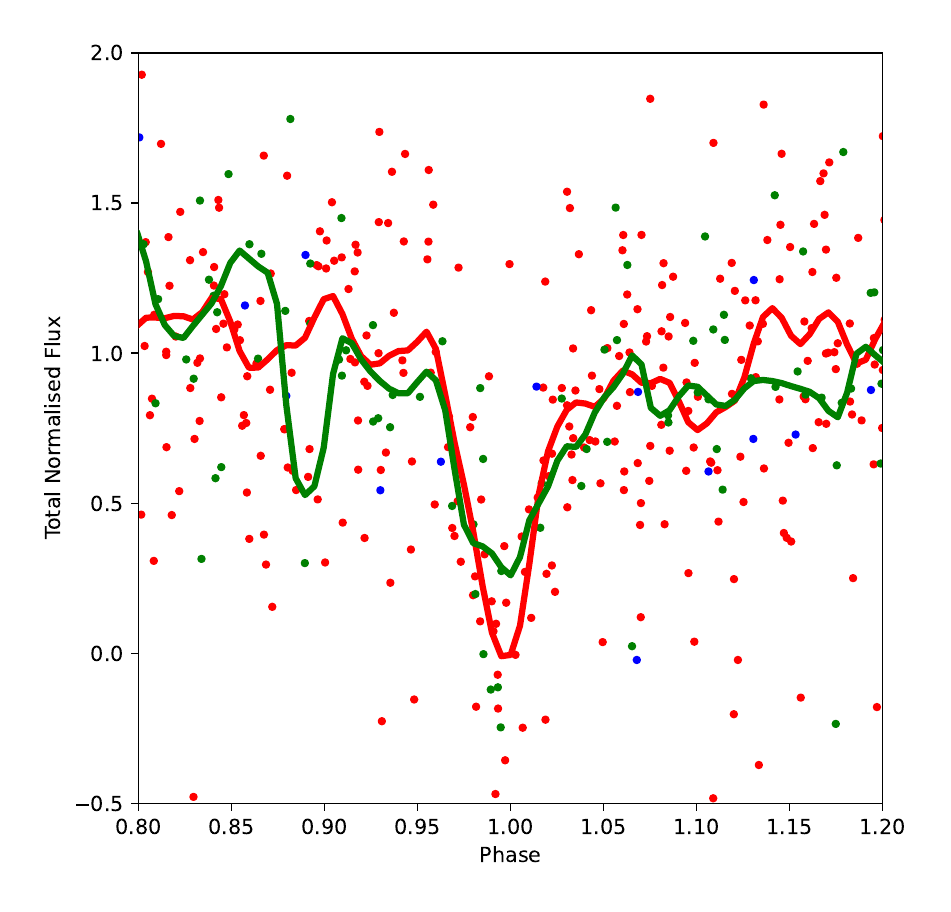}
\caption{ZTF data of ZTF20aabowdt. On the left, the upper panel shows the lightcurve, the middle panel shows the folded lightcurve ($P=25.92$ minutes), and the final panel shows the periodogram. The median uncertainty on the ZTF measurements is 33\% and is not shown for clarity. On the right is a zoom-in on the eclipse of the folded lightcurve. The smoothed lightcurve (using a Gaussian kernel) is overplotted. Legend: The data of the $g$, $r$, $i$ bands are shown in green, red, and blue}
\label{Fig:1}
\end{figure*}

\begin{figure*}[h!]
\includegraphics[width = 9 cm]{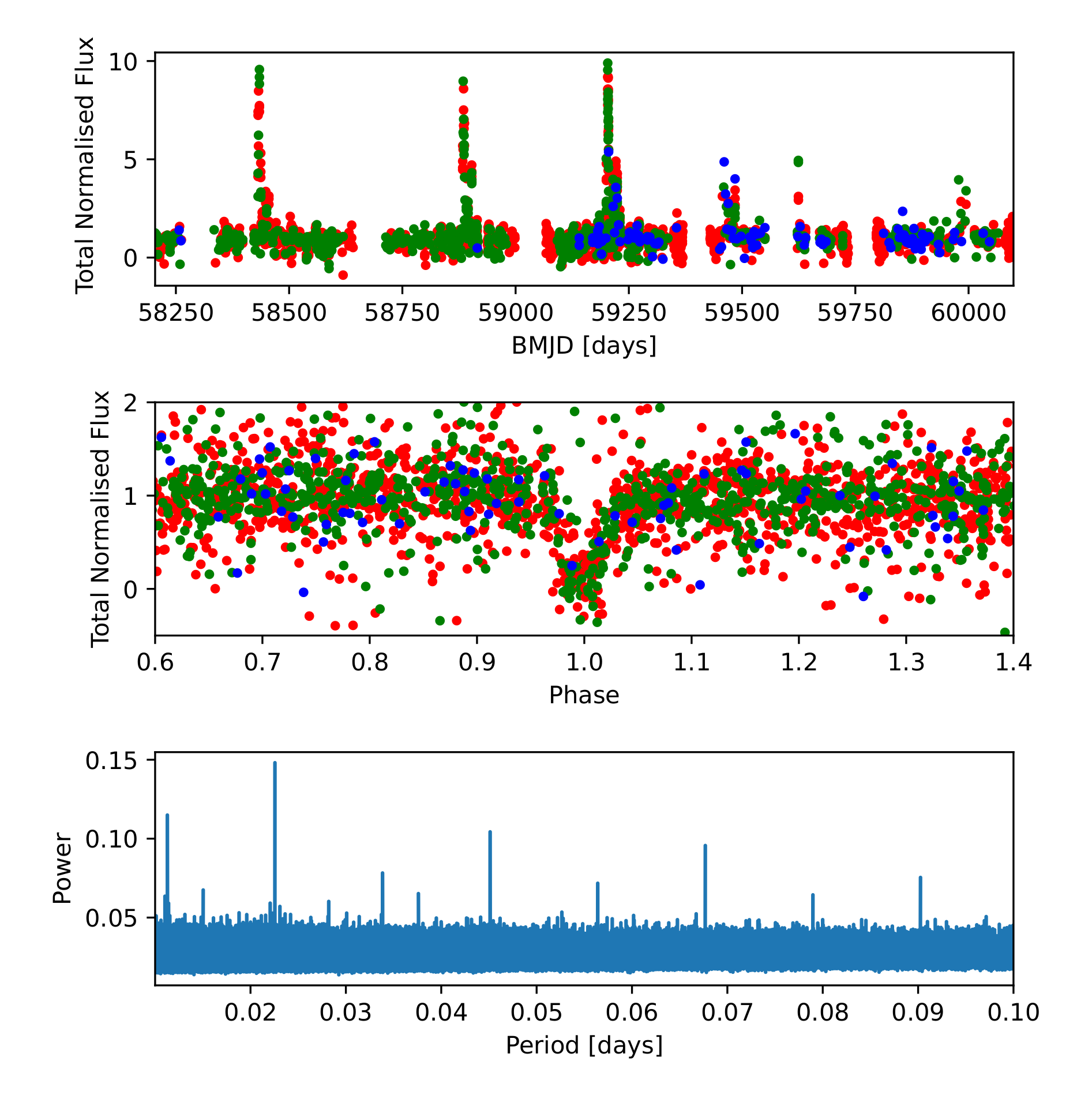}
\includegraphics[width = 9 cm]{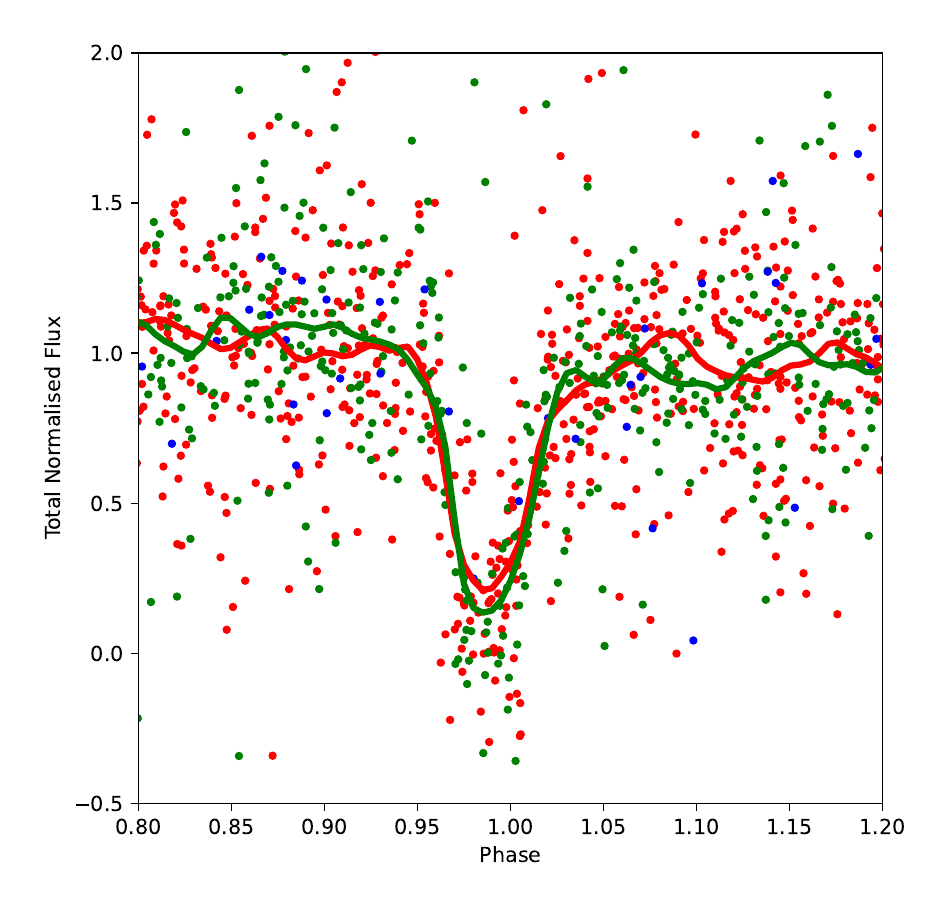}
\caption{ZTF data of ZTF18acgmwpt. On the left, upper panel is the lightcurve, middle panel is the folded lightcurve, and final panel is the periodogram. The median uncertainty is 21\% and is not shown for clarity. On the right is a zoom-in on the eclipse of the folded lightcurve ($P=32.47$ minutes). The smoothed lightcurve (using a Gaussian kernel) is overplotted. Legend: The $g$, $r$, $i$ band data is shown in green, red and blue}
\label{Fig:2}
\end{figure*}

\begin{figure*}[h!]
\includegraphics[width = 9 cm]{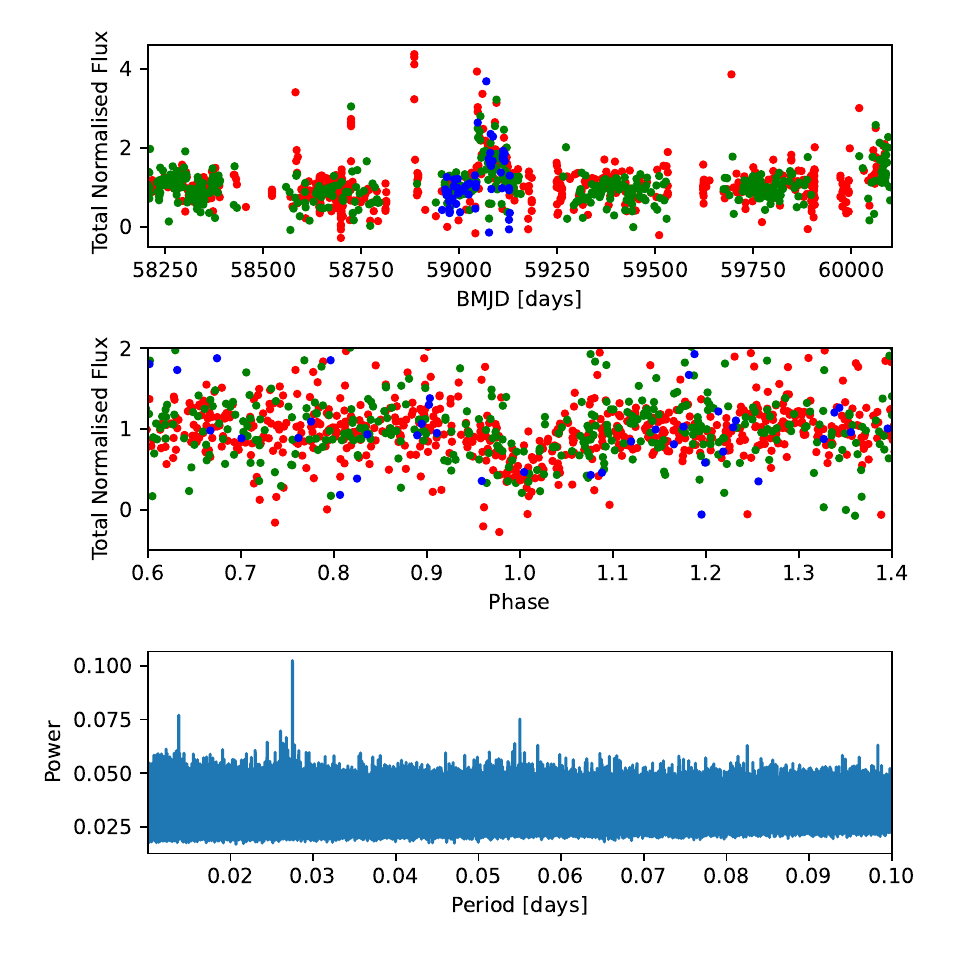}
\includegraphics[width = 9 cm]{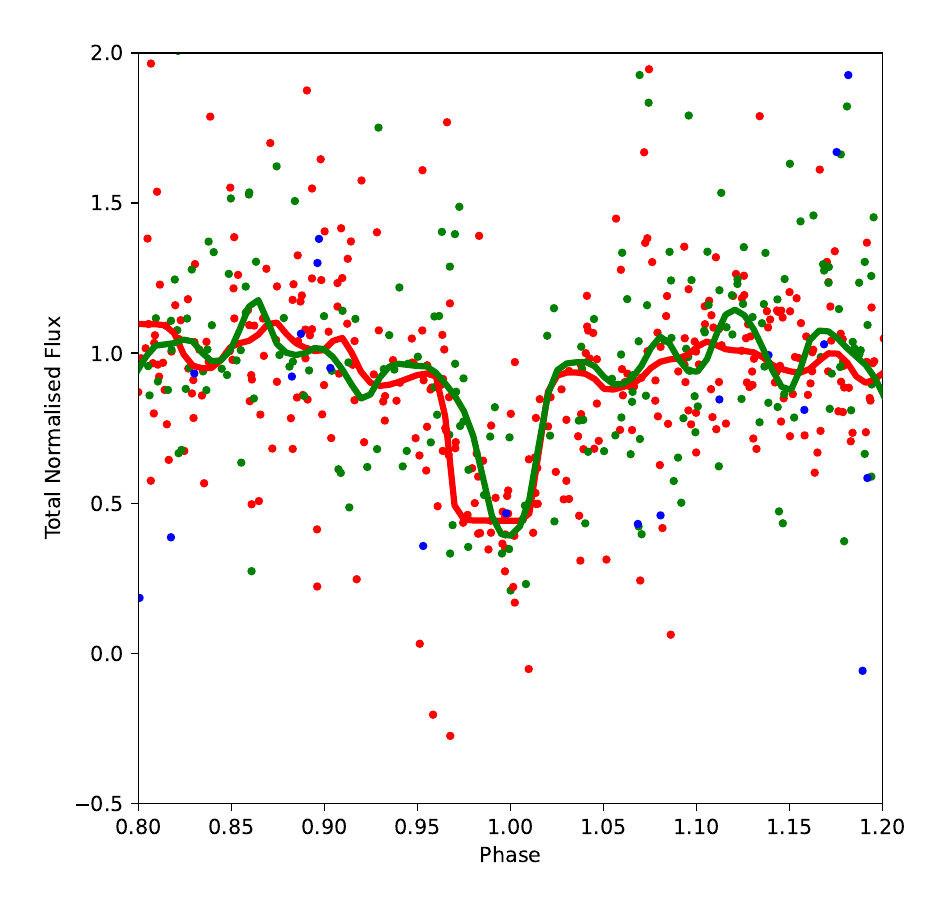}
\caption{ZTF data of ZTF19abugzba. On the left, upper panel is the lightcurve, middle panel is the folded lightcurve ($P=39.61$ minutes), and final panel is the periodogram. The median uncertainty is 11\% and is not shown for clarity. On the right is a zoom-in on the eclipse of the folded lightcurve.  The smoothed lightcurve (using a Gaussian kernel) is overplotted. Legend: The $g$, $r$, $i$ band data is shown in green, red and blue}
\label{Fig:3}
\end{figure*}

\begin{figure*}[h!]
\includegraphics[width = 9 cm]{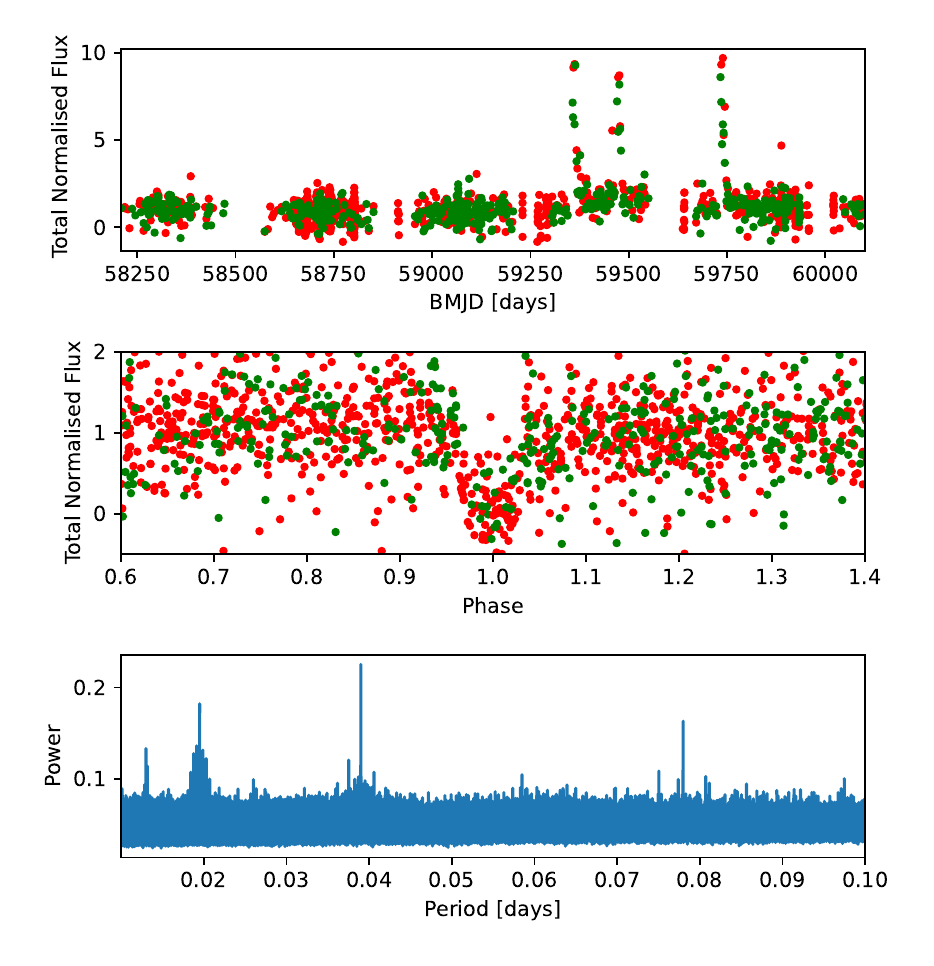}
\includegraphics[width = 9 cm]{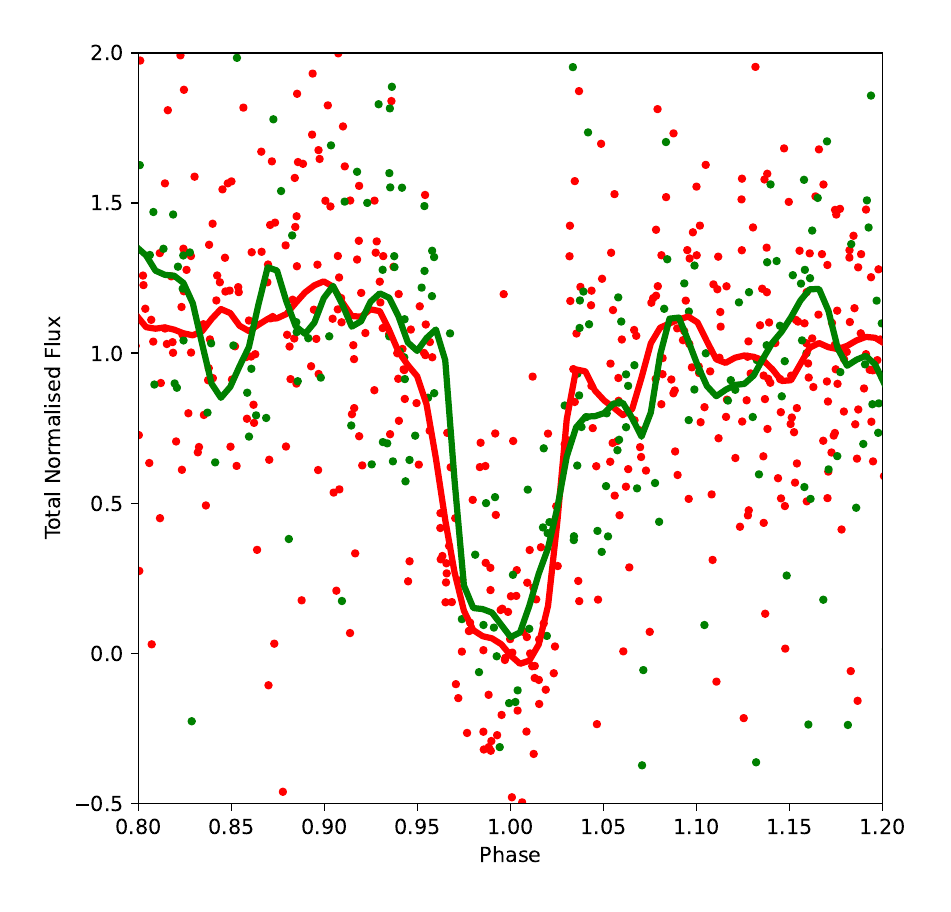}
\caption{ZTF data of ZTF21abbxnbm. On the left, upper panel is the lightcurve, middle panel is the folded lightcurve ($P=56.16$ minutes), and final panel is the periodogram. The median uncertainty is 35\% and is not shown for clarity. On the right is a zoom-in on the eclipse of the folded lightcurve. The smoothed lightcurve (using a Gaussian kernel) is overplotted. Legend: The $g$, $r$, $i$ band data is shown in green, red and blue}
\label{Fig:4}
\end{figure*}

Figures \ref{Fig:1}--\ref{Fig:4} show the ZTF lightcurves for each object. The system ZTF20aabowdt shows two super-outbursts, the system ZTF18acgmwpt shows three super-outbursts, system ZTF19abugzba shows a single super-outburst, while the system ZTF21abbxnbm shows three super-outbursts. All four systems show other outbursts with smaller amplitudes and duration.\\

ZTF20aabowdt, ZTF18acgmwpt, and ZTF21abbxnbm have deep eclipses (flux decreases to almost 0). However, ZTF19abugzba has a shallow eclipse with a depth of only $\approx$50\%. ZTF20aabowdt also seems to exhibit a narrower eclipse than other systems.

\section{Discussion}\label{sec:Discussion}
In this section, we compare the observed properties of the four systems with the currently known AM CVn systems and other accreting ultracompact binaries.
\subsection{Outburst behaviour}
If we assume that all four systems are AM CVn systems, we can estimate the expected outburst recurrence time. 
To calculate the expected recurrence time of the four systems, we used the empirical equation provided by \cite{2015MNRAS.446..391L}:
\[P_\mathrm{rec} = (1.53 \times 10^{\ -9} )P_\mathrm{orb}^{\ 7.35} + 24.7\]
with $P_\mathrm{rec}$ is the super-outburst recurrence time in days and $P_\mathrm{orb}$ is the orbital period in minutes. 
The predicted recurrence times for ZTF20aabowdt, ZTF18acgmwpt, ZTF19abugzba, and ZTF21abbxnbm in days are: 62, 221, 873 ($\approx$2.4 yrs), and 11065 ($\approx$30 yrs), respectively. 

ZTF20aabowdt shows two super-outbursts and 9 outbursts with smaller amplitudes and durations. The predicted recurrence time for this system (62 days) suggests that it should show more super-outbursts, which are likely missed due to the poor sampling by ZTF due to the system's low declination. 

ZTF18acgmwpt shows three clear super-outbursts and three with smaller amplitudes and shorter durations. The times between the three large super-outbursts are 450 and 320 days. Notably, no super-outbursts were detected in the last 2 years of data. 
The system's expected outburst recurrence time (221 days) suggests that we might be missing at least two more super-outbursts, or that the outburst recurrence time is longer compared to other AM CVn systems with this orbital period. Due to the gaps in the ZTF data, we cannot draw a definite conclusion.

ZTF19abugzba shows a single, faint, super-outburst and multiple echo outbursts with smaller amplitudes and short durations starting at 59050 MJD. Similar faint super-outbursts with many echo outbursts have been seen in other AM CVn system \citep[e.g.][]{2021PASJ...73.1375K}. The calculated recurrence time (873 days) suggests that one or two super-outbursts are missed because of lack of observation. We see increased activity at the very end of the lightcurve which suggests that a super-outburst occurred in a gap in the data. As in the previous cases, we cannot determine if we missed the outburst or if the outburst recurrence rate is lower than expected.

ZTF21abbxnbm does show a clear deviation from predictions since the system's calculated recurrence time suggests a single outburst every $\approx$30 years where we see three super-outbursts in a period of roughly three years. This behaviour is very inconsistent with the behaviour of other known long-period AM CVn. Similar AM CVn systems in \citet{2015MNRAS.446..391L} have been reported to have no visible outbursts or to have only a detected outburst. Helium CVs on the other hand do tend to show frequent outbursts at orbital periods of 50 to 70 minutes. Therefore, we speculate that ZTF21abbxnbm is a Helium CV instead of a classical long-period AM CVn system.

\subsection{Eclipse shape}
The eclipse depth for AM CVn systems (and most other accreting ultracompact binaries) is mostly determined by the brightness of the accretion disk because the donor is cold and does not contribute any significant amount of light in the optical. For long-period systems (which typically have a lower accretion rate), the accretion disk is typically fainter, resulting in a deeper eclipse \citep[e.g.][]{2006ApJ...640..466B}. YZ~Lmi, the first discovered eclipsing AM CVn \cite{2011MNRAS.410.1113C} with period of 28.31 minutes, has a white dwarf that contributes about 70\% of all light\footnote{YZ~LMi shows a grazing eclipse and therefore the eclipse is only 60\% deep. If the inclination for YZ~LMi was 90$^{\circ}$, the eclipse depth would be $\approx$97\%.}  
ZTF20aabowdt shows an eclipse depth of close to a 100\%. This means that the disk brightness is still only a small fraction of the overall luminosity of the system and is very similar to the slightly longer period eclipsing system YZ~Lmi. The fact that we are seeing a much smaller contribution from the disk compared to the prediction from \citet[]{2006ApJ...640..466B} is mostly due to the high inclination angle which means the visible projected area is very small.

ZTF18acgmwpt also shows a deep eclipse and has an orbital period of 32.47 minutes. It is comparable to ZTFJ0407\textminus0007 in period and also shows a deep eclipse \citep{2022MNRAS.512.5440V}. On the other hand, ZTF19abugzba has a shallow depth of eclipse that only reaches about 50\%, which is unexpected for a system with a period of 39.61 minutes. The shallow depth of eclipse could be due to an increase in the accretion disk activity, but, much more likely, we are seeing only partial eclipses of the white dwarf.

ZTF21abbxnbm behaves as expected with a deep eclipse comparable with its long-period of 56.16 minutes. If this system is an Helium-CV instead of a long-period AM CVn system, we would also expect to see deep eclipses because in both cases the donor stars are cold.

\section{Conclusion}\label{sec:conclusion}
We searched for periodic eclipses in a sample of dwarf novae found with ZTF. The BLS method was used to search for orbital periods between 4.32-65 minutes. Four new eclipsing accreting ultracompact binaries with orbital periods between 25.92--56.16 minutes were discovered. The shorter-period systems behave roughly as expected of AM CV systems, with one of the systems only showing a grazing eclipse. The fourth system with the longest period (\textit{P} = 56.16 minutes) is unusual because it exhibits three super-outbursts in three years, which suggests that it is a Helium CV instead of a typical long-period AM CVn. Both high-speed photometry and spectroscopic follow-up observations are underway to better characterise the four new systems.

\begin{acknowledgements}
We thank Tony Rodriguez and Paula Szkody for feedback on the manuscript.

Jasmine Khalil appreciates the support from the ASPIRE programme supervisors and organisers as well as the opportunity given by the University of Amsterdam.

This publication is part of the project "The life and death of white dwarf binary stars" (with project number VI.Veni.212.201) of the research programme NWO Talent Programme Veni Science domain 2021 which is financed by the Dutch Research Council (NWO).

Based on observations obtained with the Samuel Oschin Telescope 48-inch and the 60-inch Telescope at the Palomar Observatory as part of the Zwicky Transient Facility project. ZTF is supported by the National Science Foundation under Grants No. AST-1440341 and AST-2034437 and a collaboration including current partners Caltech, IPAC, the Weizmann Institute of Science, the Oskar Klein Center at Stockholm University, the University of Maryland, Deutsches Elektronen-Synchrotron and Humboldt University, the TANGO Consortium of Taiwan, the University of Wisconsin at Milwaukee, Trinity College Dublin, Lawrence Livermore National Laboratories, IN2P3, University of Warwick, Ruhr University Bochum, Northwestern University and former partners the University of Washington, Los Alamos National Laboratories, and Lawrence Berkeley National Laboratories. Operations are conducted by COO, IPAC, and UW.

The ZTF forced-photometry service was funded under the Heising-Simons Foundation grant \#12540303 (PI: Graham).

\end{acknowledgements}

%
%

\bibliographystyle{aa} 
\bibliography{biblio}

\begin{thebibliography}{35}
\expandafter\ifx\csname natexlab\endcsname\relax\def\natexlab#1{#1}\fi

\bibitem[{{Bailer-Jones} {et~al.}(2018){Bailer-Jones}, {Rybizki}, {Fouesneau},
  {Mantelet}, \& {Andrae}}]{2018AJ....156...58B}
{Bailer-Jones}, C.~A.~L., {Rybizki}, J., {Fouesneau}, M., {Mantelet}, G., \&
  {Andrae}, R. 2018, \aj, 156, 58

\bibitem[{{Bellm} {et~al.}(2019){Bellm}, {Kulkarni}, {Graham}, {Dekany},
  {Smith}, {Riddle}, {Masci}, {Helou}, {Prince}, {Adams}, {Barbarino},
  {Barlow}, {Bauer}, {Beck}, {Belicki}, {Biswas}, {Blagorodnova}, {Bodewits},
  {Bolin}, {Brinnel}, {Brooke}, {Bue}, {Bulla}, {Burruss}, {Cenko}, {Chang},
  {Connolly}, {Coughlin}, {Cromer}, {Cunningham}, {De}, {Delacroix}, {Desai},
  {Duev}, {Eadie}, {Farnham}, {Feeney}, {Feindt}, {Flynn}, {Franckowiak},
  {Frederick}, {Fremling}, {Gal-Yam}, {Gezari}, {Giomi}, {Goldstein},
  {Golkhou}, {Goobar}, {Groom}, {Hacopians}, {Hale}, {Henning}, {Ho}, {Hover},
  {Howell}, {Hung}, {Huppenkothen}, {Imel}, {Ip}, {Ivezi{\'c}}, {Jackson},
  {Jones}, {Juric}, {Kasliwal}, {Kaspi}, {Kaye}, {Kelley}, {Kowalski},
  {Kramer}, {Kupfer}, {Landry}, {Laher}, {Lee}, {Lin}, {Lin}, {Lunnan},
  {Giomi}, {Mahabal}, {Mao}, {Miller}, {Monkewitz}, {Murphy}, {Ngeow},
  {Nordin}, {Nugent}, {Ofek}, {Patterson}, {Penprase}, {Porter}, {Rauch},
  {Rebbapragada}, {Reiley}, {Rigault}, {Rodriguez}, {van Roestel}, {Rusholme},
  {van Santen}, {Schulze}, {Shupe}, {Singer}, {Soumagnac}, {Stein}, {Surace},
  {Sollerman}, {Szkody}, {Taddia}, {Terek}, {Van Sistine}, {van Velzen},
  {Vestrand}, {Walters}, {Ward}, {Ye}, {Yu}, {Yan}, \&
  {Zolkower}}]{2019PASP..131a8002B}
{Bellm}, E.~C., {Kulkarni}, S.~R., {Graham}, M.~J., {et~al.} 2019, \pasp, 131,
  018002

\bibitem[{Bhatti {et~al.}(2017)Bhatti, Bouma, \& Wallace}]{wbhatti_astrobase}
Bhatti, W., Bouma, L.~G., \& Wallace, J. 2017, \texttt{Astrobase}

\bibitem[{{Bildsten} {et~al.}(2006){Bildsten}, {Townsley}, {Deloye}, \&
  {Nelemans}}]{2006ApJ...640..466B}
{Bildsten}, L., {Townsley}, D.~M., {Deloye}, C.~J., \& {Nelemans}, G. 2006,
  \apj, 640, 466

\bibitem[{{Breedt} {et~al.}(2012){Breedt}, {G{\"a}nsicke}, {Marsh}, {Steeghs},
  {Drake}, \& {Copperwheat}}]{2012MNRAS.425.2548B}
{Breedt}, E., {G{\"a}nsicke}, B.~T., {Marsh}, T.~R., {et~al.} 2012, \mnras,
  425, 2548

\bibitem[{{Burdge} {et~al.}(2022){Burdge}, {El-Badry}, {Marsh}, {Rappaport},
  {Brown}, {Caiazzo}, {Chakrabarty}, {Dhillon}, {Fuller}, {G{\"a}nsicke},
  {Graham}, {Kara}, {Kulkarni}, {Littlefair}, {Mr{\'o}z}, {Rodr{\'\i}guez-Gil},
  {Roestel}, {Simcoe}, {Bellm}, {Drake}, {Dekany}, {Groom}, {Laher}, {Masci},
  {Riddle}, {Smith}, \& {Prince}}]{2022Natur.610..467B}
{Burdge}, K.~B., {El-Badry}, K., {Marsh}, T.~R., {et~al.} 2022, \nat, 610, 467

\bibitem[{{Burdge} {et~al.}(2020){Burdge}, {Prince}, {Fuller}, {Kaplan},
  {Marsh}, {Tremblay}, {Zhuang}, {Bellm}, {Caiazzo}, {Coughlin}, {Dhillon},
  {Gaensicke}, {Rodr{\'\i}guez-Gil}, {Graham}, {Hermes}, {Kupfer},
  {Littlefair}, {Mr{\'o}z}, {Phinney}, {van Roestel}, {Yao}, {Dekany}, {Drake},
  {Duev}, {Hale}, {Feeney}, {Helou}, {Kaye}, {Mahabal}, {Masci}, {Riddle},
  {Smith}, {Soumagnac}, \& {Kulkarni}}]{2020ApJ...905...32B}
{Burdge}, K.~B., {Prince}, T.~A., {Fuller}, J., {et~al.} 2020, \apj, 905, 32

\bibitem[{{Chambers} {et~al.}(2016){Chambers}, {Magnier}, {Metcalfe},
  {Flewelling}, {Huber}, {Waters}, {Denneau}, {Draper}, {Farrow}, {Finkbeiner},
  {Holmberg}, {Koppenhoefer}, {Price}, {Rest}, {Saglia}, {Schlafly}, {Smartt},
  {Sweeney}, {Wainscoat}, {Burgett}, {Chastel}, {Grav}, {Heasley}, {Hodapp},
  {Jedicke}, {Kaiser}, {Kudritzki}, {Luppino}, {Lupton}, {Monet}, {Morgan},
  {Onaka}, {Shiao}, {Stubbs}, {Tonry}, {White}, {Ba{\~n}ados}, {Bell},
  {Bender}, {Bernard}, {Boegner}, {Boffi}, {Botticella}, {Calamida},
  {Casertano}, {Chen}, {Chen}, {Cole}, {Deacon}, {Frenk}, {Fitzsimmons},
  {Gezari}, {Gibbs}, {Goessl}, {Goggia}, {Gourgue}, {Goldman}, {Grant},
  {Grebel}, {Hambly}, {Hasinger}, {Heavens}, {Heckman}, {Henderson}, {Henning},
  {Holman}, {Hopp}, {Ip}, {Isani}, {Jackson}, {Keyes}, {Koekemoer}, {Kotak},
  {Le}, {Liska}, {Long}, {Lucey}, {Liu}, {Martin}, {Masci}, {McLean}, {Mindel},
  {Misra}, {Morganson}, {Murphy}, {Obaika}, {Narayan}, {Nieto-Santisteban},
  {Norberg}, {Peacock}, {Pier}, {Postman}, {Primak}, {Rae}, {Rai}, {Riess},
  {Riffeser}, {Rix}, {R{\"o}ser}, {Russel}, {Rutz}, {Schilbach}, {Schultz},
  {Scolnic}, {Strolger}, {Szalay}, {Seitz}, {Small}, {Smith}, {Soderblom},
  {Taylor}, {Thomson}, {Taylor}, {Thakar}, {Thiel}, {Thilker}, {Unger},
  {Urata}, {Valenti}, {Wagner}, {Walder}, {Walter}, {Watters}, {Werner},
  {Wood-Vasey}, \& {Wyse}}]{2016arXiv161205560C}
{Chambers}, K.~C., {Magnier}, E.~A., {Metcalfe}, N., {et~al.} 2016, arXiv
  e-prints, arXiv:1612.05560

\bibitem[{{Copperwheat} {et~al.}(2011){Copperwheat}, {Marsh}, {Littlefair},
  {Dhillon}, {Ramsay}, {Drake}, {G{\"a}nsicke}, {Groot}, {Hakala}, {Koester},
  {Nelemans}, {Roelofs}, {Southworth}, {Steeghs}, \&
  {Tulloch}}]{2011MNRAS.410.1113C}
{Copperwheat}, C.~M., {Marsh}, T.~R., {Littlefair}, S.~P., {et~al.} 2011,
  \mnras, 410, 1113

\bibitem[{{Coughlin} {et~al.}(2023){Coughlin}, {Bloom}, {Nir}, {Antier}, {du
  Laz}, {van der Walt}, {Crellin-Quick}, {Culino}, {Duev}, {Goldstein},
  {Healy}, {Karambelkar}, {Lilleboe}, {Shin}, {Singer}, {Ahumada}, {Anand},
  {Bellm}, {Dekany}, {Graham}, {Kasliwal}, {Kostadinova}, {Kiendrebeogo},
  {Kulkarni}, {Jenkins}, {LeBaron}, {Mahabal}, {Neill}, {Parazin}, {Peloton},
  {Perley}, {Riddle}, {Rusholme}, {van Santen}, {Sollerman}, {Stein}, {Turpin},
  {Wold}, {Amat}, {Bonnefon}, {Bonnefoy}, {Flament}, {Kerkow}, {Kishore},
  {Jani}, {Mahanty}, {Liu}, {Llinares}, {Makarison}, {Olli{\'e}ric}, {Perez},
  {Pont}, \& {Sharma}}]{coughlin2023}
{Coughlin}, M.~W., {Bloom}, J.~S., {Nir}, G., {et~al.} 2023, \apjs, 267, 31

\bibitem[{{Dekany} {et~al.}(2020){Dekany}, {Smith}, {Riddle}, {Feeney},
  {Porter}, {Hale}, {Zolkower}, {Belicki}, {Kaye}, {Henning}, {Walters},
  {Cromer}, {Delacroix}, {Rodriguez}, {Reiley}, {Mao}, {Hover}, {Murphy},
  {Burruss}, {Baker}, {Kowalski}, {Reif}, {Mueller}, {Bellm}, {Graham}, \&
  {Kulkarni}}]{2020PASP..132c8001D}
{Dekany}, R., {Smith}, R.~M., {Riddle}, R., {et~al.} 2020, \pasp, 132, 038001

\bibitem[{{Gaia Collaboration} {et~al.}(2016){Gaia Collaboration}, {Prusti},
  {de Bruijne}, {Brown}, {Vallenari}, {Babusiaux}, {Bailer-Jones}, {Bastian},
  {Biermann}, {Evans}, {Eyer}, {Jansen}, {Jordi}, {Klioner}, {Lammers},
  {Lindegren}, {Luri}, {Mignard}, {Milligan}, {Panem}, {Poinsignon},
  {Pourbaix}, {Randich}, {Sarri}, {Sartoretti}, {Siddiqui}, {Soubiran},
  {Valette}, {van Leeuwen}, {Walton}, {Aerts}, {Arenou}, {Cropper}, {Drimmel},
  {H{\o}g}, {Katz}, {Lattanzi}, {O'Mullane}, {Grebel}, {Holland}, {Huc},
  {Passot}, {Bramante}, {Cacciari}, {Casta{\~n}eda}, {Chaoul}, {Cheek}, {De
  Angeli}, {Fabricius}, {Guerra}, {Hern{\'a}ndez}, {Jean-Antoine-Piccolo},
  {Masana}, {Messineo}, {Mowlavi}, {Nienartowicz}, {Ord{\'o}{\~n}ez-Blanco},
  {Panuzzo}, {Portell}, {Richards}, {Riello}, {Seabroke}, {Tanga},
  {Th{\'e}venin}, {Torra}, {Els}, {Gracia-Abril}, {Comoretto},
  {Garcia-Reinaldos}, {Lock}, {Mercier}, {Altmann}, {Andrae}, {Astraatmadja},
  {Bellas-Velidis}, {Benson}, {Berthier}, {Blomme}, {Busso}, {Carry},
  {Cellino}, {Clementini}, {Cowell}, {Creevey}, {Cuypers}, {Davidson}, {De
  Ridder}, {de Torres}, {Delchambre}, {Dell'Oro}, {Ducourant}, {Fr{\'e}mat},
  {Garc{\'\i}a-Torres}, {Gosset}, {Halbwachs}, {Hambly}, {Harrison}, {Hauser},
  {Hestroffer}, {Hodgkin}, {Huckle}, {Hutton}, {Jasniewicz}, {Jordan},
  {Kontizas}, {Korn}, {Lanzafame}, {Manteiga}, {Moitinho}, {Muinonen},
  {Osinde}, {Pancino}, {Pauwels}, {Petit}, {Recio-Blanco}, {Robin}, {Sarro},
  {Siopis}, {Smith}, {Smith}, {Sozzetti}, {Thuillot}, {van Reeven}, {Viala},
  {Abbas}, {Abreu Aramburu}, {Accart}, {Aguado}, {Allan}, {Allasia},
  {Altavilla}, {{\'A}lvarez}, {Alves}, {Anderson}, {Andrei}, {Anglada Varela},
  {Antiche}, {Antoja}, {Ant{\'o}n}, {Arcay}, {Atzei}, {Ayache}, {Bach},
  {Baker}, {Balaguer-N{\'u}{\~n}ez}, {Barache}, {Barata}, {Barbier}, {Barblan},
  {Baroni}, {Barrado y Navascu{\'e}s}, {Barros}, {Barstow}, {Becciani},
  {Bellazzini}, {Bellei}, {Bello Garc{\'\i}a}, {Belokurov}, {Bendjoya},
  {Berihuete}, {Bianchi}, {Bienaym{\'e}}, {Billebaud}, {Blagorodnova},
  {Blanco-Cuaresma}, {Boch}, {Bombrun}, {Borrachero}, {Bouquillon}, {Bourda},
  {Bouy}, {Bragaglia}, {Breddels}, {Brouillet}, {Br{\"u}semeister},
  {Bucciarelli}, {Budnik}, {Burgess}, {Burgon}, {Burlacu}, {Busonero}, {Buzzi},
  {Caffau}, {Cambras}, {Campbell}, {Cancelliere}, {Cantat-Gaudin}, {Carlucci},
  {Carrasco}, {Castellani}, {Charlot}, {Charnas}, {Charvet}, {Chassat},
  {Chiavassa}, {Clotet}, {Cocozza}, {Collins}, {Collins}, {Costigan}, {Crifo},
  {Cross}, {Crosta}, {Crowley}, {Dafonte}, {Damerdji}, {Dapergolas}, {David},
  {David}, {De Cat}, {de Felice}, {de Laverny}, {De Luise}, {De March}, {de
  Martino}, {de Souza}, {Debosscher}, {del Pozo}, {Delbo}, {Delgado},
  {Delgado}, {di Marco}, {Di Matteo}, {Diakite}, {Distefano}, {Dolding}, {Dos
  Anjos}, {Drazinos}, {Dur{\'a}n}, {Dzigan}, {Ecale}, {Edvardsson}, {Enke},
  {Erdmann}, {Escolar}, {Espina}, {Evans}, {Eynard Bontemps}, {Fabre},
  {Fabrizio}, {Faigler}, {Falc{\~a}o}, {Farr{\`a}s Casas}, {Faye}, {Federici},
  {Fedorets}, {Fern{\'a}ndez-Hern{\'a}ndez}, {Fernique}, {Fienga}, {Figueras},
  {Filippi}, {Findeisen}, {Fonti}, {Fouesneau}, {Fraile}, {Fraser}, {Fuchs},
  {Furnell}, {Gai}, {Galleti}, {Galluccio}, {Garabato}, {Garc{\'\i}a-Sedano},
  {Gar{\'e}}, {Garofalo}, {Garralda}, {Gavras}, {Gerssen}, {Geyer}, {Gilmore},
  {Girona}, {Giuffrida}, {Gomes}, {Gonz{\'a}lez-Marcos},
  {Gonz{\'a}lez-N{\'u}{\~n}ez}, {Gonz{\'a}lez-Vidal}, {Granvik}, {Guerrier},
  {Guillout}, {Guiraud}, {G{\'u}rpide}, {Guti{\'e}rrez-S{\'a}nchez}, {Guy},
  {Haigron}, {Hatzidimitriou}, {Haywood}, {Heiter}, {Helmi}, {Hobbs},
  {Hofmann}, {Holl}, {Holland}, {Hunt}, {Hypki}, {Icardi}, {Irwin}, {Jevardat
  de Fombelle}, {Jofr{\'e}}, {Jonker}, {Jorissen}, {Julbe}, {Karampelas},
  {Kochoska}, {Kohley}, {Kolenberg}, {Kontizas}, {Koposov}, {Kordopatis},
  {Koubsky}, {Kowalczyk}, {Krone-Martins}, {Kudryashova}, {Kull}, {Bachchan},
  {Lacoste-Seris}, {Lanza}, {Lavigne}, {Le Poncin-Lafitte}, {Lebreton},
  {Lebzelter}, {Leccia}, {Leclerc}, {Lecoeur-Taibi}, {Lemaitre}, {Lenhardt},
  {Leroux}, {Liao}, {Licata}, {Lindstr{\o}m}, {Lister}, {Livanou}, {Lobel},
  {L{\"o}ffler}, {L{\'o}pez}, {Lopez-Lozano}, {Lorenz}, {Loureiro},
  {MacDonald}, {Magalh{\~a}es Fernandes}, {Managau}, {Mann}, {Mantelet},
  {Marchal}, {Marchant}, {Marconi}, {Marie}, {Marinoni}, {Marrese},
  {Marschalk{\'o}}, {Marshall}, {Mart{\'\i}n-Fleitas}, {Martino}, {Mary},
  {Matijevi{\v{c}}}, {Mazeh}, {McMillan}, {Messina}, {Mestre}, {Michalik},
  {Millar}, {Miranda}, {Molina}, {Molinaro}, {Molinaro}, {Moln{\'a}r},
  {Moniez}, {Montegriffo}, {Monteiro}, {Mor}, {Mora}, {Morbidelli}, {Morel},
  {Morgenthaler}, {Morley}, {Morris}, {Mulone}, {Muraveva}, {Musella},
  {Narbonne}, {Nelemans}, {Nicastro}, {Noval}, {Ord{\'e}novic},
  {Ordieres-Mer{\'e}}, {Osborne}, {Pagani}, {Pagano}, {Pailler}, {Palacin},
  {Palaversa}, {Parsons}, {Paulsen}, {Pecoraro}, {Pedrosa}, {Pentik{\"a}inen},
  {Pereira}, {Pichon}, {Piersimoni}, {Pineau}, {Plachy}, {Plum}, {Poujoulet},
  {Pr{\v{s}}a}, {Pulone}, {Ragaini}, {Rago}, {Rambaux}, {Ramos-Lerate},
  {Ranalli}, {Rauw}, {Read}, {Regibo}, {Renk}, {Reyl{\'e}}, {Ribeiro},
  {Rimoldini}, {Ripepi}, {Riva}, {Rixon}, {Roelens}, {Romero-G{\'o}mez},
  {Rowell}, {Royer}, {Rudolph}, {Ruiz-Dern}, {Sadowski}, {Sagrist{\`a}
  Sell{\'e}s}, {Sahlmann}, {Salgado}, {Salguero}, {Sarasso}, {Savietto},
  {Schnorhk}, {Schultheis}, {Sciacca}, {Segol}, {Segovia}, {Segransan},
  {Serpell}, {Shih}, {Smareglia}, {Smart}, {Smith}, {Solano}, {Solitro},
  {Sordo}, {Soria Nieto}, {Souchay}, {Spagna}, {Spoto}, {Stampa}, {Steele},
  {Steidelm{\"u}ller}, {Stephenson}, {Stoev}, {Suess}, {S{\"u}veges}, {Surdej},
  {Szabados}, {Szegedi-Elek}, {Tapiador}, {Taris}, {Tauran}, {Taylor},
  {Teixeira}, {Terrett}, {Tingley}, {Trager}, {Turon}, {Ulla}, {Utrilla},
  {Valentini}, {van Elteren}, {Van Hemelryck}, {van Leeuwen}, {Varadi},
  {Vecchiato}, {Veljanoski}, {Via}, {Vicente}, {Vogt}, {Voss}, {Votruba},
  {Voutsinas}, {Walmsley}, {Weiler}, {Weingrill}, {Werner}, {Wevers},
  {Whitehead}, {Wyrzykowski}, {Yoldas}, {{\v{Z}}erjal}, {Zucker}, {Zurbach},
  {Zwitter}, {Alecu}, {Allen}, {Allende Prieto}, {Amorim},
  {Anglada-Escud{\'e}}, {Arsenijevic}, {Azaz}, {Balm}, {Beck}, {Bernstein},
  {Bigot}, {Bijaoui}, {Blasco}, {Bonfigli}, {Bono}, {Boudreault}, {Bressan},
  {Brown}, {Brunet}, {Bunclark}, {Buonanno}, {Butkevich}, {Carret}, {Carrion},
  {Chemin}, {Ch{\'e}reau}, {Corcione}, {Darmigny}, {de Boer}, {de Teodoro}, {de
  Zeeuw}, {Delle Luche}, {Domingues}, {Dubath}, {Fodor}, {Fr{\'e}zouls},
  {Fries}, {Fustes}, {Fyfe}, {Gallardo}, {Gallegos}, {Gardiol}, {Gebran},
  {Gomboc}, {G{\'o}mez}, {Grux}, {Gueguen}, {Heyrovsky}, {Hoar}, {Iannicola},
  {Isasi Parache}, {Janotto}, {Joliet}, {Jonckheere}, {Keil}, {Kim},
  {Klagyivik}, {Klar}, {Knude}, {Kochukhov}, {Kolka}, {Kos}, {Kutka}, {Lainey},
  {LeBouquin}, {Liu}, {Loreggia}, {Makarov}, {Marseille}, {Martayan},
  {Martinez-Rubi}, {Massart}, {Meynadier}, {Mignot}, {Munari}, {Nguyen},
  {Nordlander}, {Ocvirk}, {O'Flaherty}, {Olias Sanz}, {Ortiz}, {Osorio},
  {Oszkiewicz}, {Ouzounis}, {Palmer}, {Park}, {Pasquato}, {Peltzer}, {Peralta},
  {P{\'e}turaud}, {Pieniluoma}, {Pigozzi}, {Poels}, {Prat}, {Prod'homme},
  {Raison}, {Rebordao}, {Risquez}, {Rocca-Volmerange}, {Rosen}, {Ruiz-Fuertes},
  {Russo}, {Sembay}, {Serraller Vizcaino}, {Short}, {Siebert}, {Silva},
  {Sinachopoulos}, {Slezak}, {Soffel}, {Sosnowska}, {Strai{\v{z}}ys}, {ter
  Linden}, {Terrell}, {Theil}, {Tiede}, {Troisi}, {Tsalmantza}, {Tur},
  {Vaccari}, {Vachier}, {Valles}, {Van Hamme}, {Veltz}, {Virtanen}, {Wallut},
  {Wichmann}, {Wilkinson}, {Ziaeepour}, \& {Zschocke}}]{2016A&A...595A...1G}
{Gaia Collaboration}, {Prusti}, T., {de Bruijne}, J.~H.~J., {et~al.} 2016,
  \aap, 595, A1

\bibitem[{{Gaia Collaboration} {et~al.}(2023){Gaia Collaboration}, {Vallenari},
  {Brown}, {Prusti}, {de Bruijne}, {Arenou}, {Babusiaux}, {Biermann},
  {Creevey}, {Ducourant}, {Evans}, {Eyer}, {Guerra}, {Hutton}, {Jordi},
  {Klioner}, {Lammers}, {Lindegren}, {Luri}, {Mignard}, {Panem}, {Pourbaix},
  {Randich}, {Sartoretti}, {Soubiran}, {Tanga}, {Walton}, {Bailer-Jones},
  {Bastian}, {Drimmel}, {Jansen}, {Katz}, {Lattanzi}, {van Leeuwen}, {Bakker},
  {Cacciari}, {Casta{\~n}eda}, {De Angeli}, {Fabricius}, {Fouesneau},
  {Fr{\'e}mat}, {Galluccio}, {Guerrier}, {Heiter}, {Masana}, {Messineo},
  {Mowlavi}, {Nicolas}, {Nienartowicz}, {Pailler}, {Panuzzo}, {Riclet}, {Roux},
  {Seabroke}, {Sordo}, {Th{\'e}venin}, {Gracia-Abril}, {Portell}, {Teyssier},
  {Altmann}, {Andrae}, {Audard}, {Bellas-Velidis}, {Benson}, {Berthier},
  {Blomme}, {Burgess}, {Busonero}, {Busso}, {C{\'a}novas}, {Carry}, {Cellino},
  {Cheek}, {Clementini}, {Damerdji}, {Davidson}, {de Teodoro}, {Nu{\~n}ez
  Campos}, {Delchambre}, {Dell'Oro}, {Esquej}, {Fern{\'a}ndez-Hern{\'a}ndez},
  {Fraile}, {Garabato}, {Garc{\'\i}a-Lario}, {Gosset}, {Haigron}, {Halbwachs},
  {Hambly}, {Harrison}, {Hern{\'a}ndez}, {Hestroffer}, {Hodgkin}, {Holl},
  {Jan{\ss}en}, {Jevardat de Fombelle}, {Jordan}, {Krone-Martins}, {Lanzafame},
  {L{\"o}ffler}, {Marchal}, {Marrese}, {Moitinho}, {Muinonen}, {Osborne},
  {Pancino}, {Pauwels}, {Recio-Blanco}, {Reyl{\'e}}, {Riello}, {Rimoldini},
  {Roegiers}, {Rybizki}, {Sarro}, {Siopis}, {Smith}, {Sozzetti}, {Utrilla},
  {van Leeuwen}, {Abbas}, {{\'A}brah{\'a}m}, {Abreu Aramburu}, {Aerts},
  {Aguado}, {Ajaj}, {Aldea-Montero}, {Altavilla}, {{\'A}lvarez}, {Alves},
  {Anders}, {Anderson}, {Anglada Varela}, {Antoja}, {Baines}, {Baker},
  {Balaguer-N{\'u}{\~n}ez}, {Balbinot}, {Balog}, {Barache}, {Barbato},
  {Barros}, {Barstow}, {Bartolom{\'e}}, {Bassilana}, {Bauchet}, {Becciani},
  {Bellazzini}, {Berihuete}, {Bernet}, {Bertone}, {Bianchi}, {Binnenfeld},
  {Blanco-Cuaresma}, {Blazere}, {Boch}, {Bombrun}, {Bossini}, {Bouquillon},
  {Bragaglia}, {Bramante}, {Breedt}, {Bressan}, {Brouillet}, {Brugaletta},
  {Bucciarelli}, {Burlacu}, {Butkevich}, {Buzzi}, {Caffau}, {Cancelliere},
  {Cantat-Gaudin}, {Carballo}, {Carlucci}, {Carnerero}, {Carrasco},
  {Casamiquela}, {Castellani}, {Castro-Ginard}, {Chaoul}, {Charlot}, {Chemin},
  {Chiaramida}, {Chiavassa}, {Chornay}, {Comoretto}, {Contursi}, {Cooper},
  {Cornez}, {Cowell}, {Crifo}, {Cropper}, {Crosta}, {Crowley}, {Dafonte},
  {Dapergolas}, {David}, {David}, {de Laverny}, {De Luise}, {De March}, {De
  Ridder}, {de Souza}, {de Torres}, {del Peloso}, {del Pozo}, {Delbo},
  {Delgado}, {Delisle}, {Demouchy}, {Dharmawardena}, {Di Matteo}, {Diakite},
  {Diener}, {Distefano}, {Dolding}, {Edvardsson}, {Enke}, {Fabre}, {Fabrizio},
  {Faigler}, {Fedorets}, {Fernique}, {Fienga}, {Figueras}, {Fournier},
  {Fouron}, {Fragkoudi}, {Gai}, {Garcia-Gutierrez}, {Garcia-Reinaldos},
  {Garc{\'\i}a-Torres}, {Garofalo}, {Gavel}, {Gavras}, {Gerlach}, {Geyer},
  {Giacobbe}, {Gilmore}, {Girona}, {Giuffrida}, {Gomel}, {Gomez},
  {Gonz{\'a}lez-N{\'u}{\~n}ez}, {Gonz{\'a}lez-Santamar{\'\i}a},
  {Gonz{\'a}lez-Vidal}, {Granvik}, {Guillout}, {Guiraud},
  {Guti{\'e}rrez-S{\'a}nchez}, {Guy}, {Hatzidimitriou}, {Hauser}, {Haywood},
  {Helmer}, {Helmi}, {Sarmiento}, {Hidalgo}, {Hilger}, {H{\l}adczuk}, {Hobbs},
  {Holland}, {Huckle}, {Jardine}, {Jasniewicz}, {Jean-Antoine Piccolo},
  {Jim{\'e}nez-Arranz}, {Jorissen}, {Juaristi Campillo}, {Julbe}, {Karbevska},
  {Kervella}, {Khanna}, {Kontizas}, {Kordopatis}, {Korn}, {K{\'o}sp{\'a}l},
  {Kostrzewa-Rutkowska}, {Kruszy{\'n}ska}, {Kun}, {Laizeau}, {Lambert},
  {Lanza}, {Lasne}, {Le Campion}, {Lebreton}, {Lebzelter}, {Leccia}, {Leclerc},
  {Lecoeur-Taibi}, {Liao}, {Licata}, {Lindstr{\o}m}, {Lister}, {Livanou},
  {Lobel}, {Lorca}, {Loup}, {Madrero Pardo}, {Magdaleno Romeo}, {Managau},
  {Mann}, {Manteiga}, {Marchant}, {Marconi}, {Marcos}, {Marcos Santos},
  {Mar{\'\i}n Pina}, {Marinoni}, {Marocco}, {Marshall}, {Martin Polo},
  {Mart{\'\i}n-Fleitas}, {Marton}, {Mary}, {Masip}, {Massari},
  {Mastrobuono-Battisti}, {Mazeh}, {McMillan}, {Messina}, {Michalik}, {Millar},
  {Mints}, {Molina}, {Molinaro}, {Moln{\'a}r}, {Monari}, {Mongui{\'o}},
  {Montegriffo}, {Montero}, {Mor}, {Mora}, {Morbidelli}, {Morel}, {Morris},
  {Muraveva}, {Murphy}, {Musella}, {Nagy}, {Noval}, {Oca{\~n}a}, {Ogden},
  {Ordenovic}, {Osinde}, {Pagani}, {Pagano}, {Palaversa}, {Palicio},
  {Pallas-Quintela}, {Panahi}, {Payne-Wardenaar}, {Pe{\~n}alosa Esteller},
  {Penttil{\"a}}, {Pichon}, {Piersimoni}, {Pineau}, {Plachy}, {Plum}, {Poggio},
  {Pr{\v{s}}a}, {Pulone}, {Racero}, {Ragaini}, {Rainer}, {Raiteri}, {Rambaux},
  {Ramos}, {Ramos-Lerate}, {Re Fiorentin}, {Regibo}, {Richards}, {Rios Diaz},
  {Ripepi}, {Riva}, {Rix}, {Rixon}, {Robichon}, {Robin}, {Robin}, {Roelens},
  {Rogues}, {Rohrbasser}, {Romero-G{\'o}mez}, {Rowell}, {Royer}, {Ruz Mieres},
  {Rybicki}, {Sadowski}, {S{\'a}ez N{\'u}{\~n}ez}, {Sagrist{\`a} Sell{\'e}s},
  {Sahlmann}, {Salguero}, {Samaras}, {Sanchez Gimenez}, {Sanna},
  {Santove{\~n}a}, {Sarasso}, {Schultheis}, {Sciacca}, {Segol}, {Segovia},
  {S{\'e}gransan}, {Semeux}, {Shahaf}, {Siddiqui}, {Siebert}, {Siltala},
  {Silvelo}, {Slezak}, {Slezak}, {Smart}, {Snaith}, {Solano}, {Solitro},
  {Souami}, {Souchay}, {Spagna}, {Spina}, {Spoto}, {Steele},
  {Steidelm{\"u}ller}, {Stephenson}, {S{\"u}veges}, {Surdej}, {Szabados},
  {Szegedi-Elek}, {Taris}, {Taylor}, {Teixeira}, {Tolomei}, {Tonello}, {Torra},
  {Torra}, {Torralba Elipe}, {Trabucchi}, {Tsounis}, {Turon}, {Ulla}, {Unger},
  {Vaillant}, {van Dillen}, {van Reeven}, {Vanel}, {Vecchiato}, {Viala},
  {Vicente}, {Voutsinas}, {Weiler}, {Wevers}, {Wyrzykowski}, {Yoldas}, {Yvard},
  {Zhao}, {Zorec}, {Zucker}, \& {Zwitter}}]{2023A&A...674A...1G}
{Gaia Collaboration}, {Vallenari}, A., {Brown}, A.~G.~A., {et~al.} 2023, \aap,
  674, A1

\bibitem[{{Garnavich} {et~al.}(2012){Garnavich}, {Littlefield}, {Marion},
  {Irwin}, {Kirshner}, \& {Vinko}}]{2012ATel.4112....1G}
{Garnavich}, P., {Littlefield}, C., {Marion}, G.~H., {et~al.} 2012, The
  Astronomer's Telegram, 4112, 1

\bibitem[{{Gentile Fusillo} {et~al.}(2021){Gentile Fusillo}, {Tremblay},
  {Cukanovaite}, {Vorontseva}, {Lallement}, {Hollands}, {G{\"a}nsicke},
  {Burdge}, {McCleery}, \& {Jordan}}]{2021MNRAS.508.3877G}
{Gentile Fusillo}, N.~P., {Tremblay}, P.~E., {Cukanovaite}, E., {et~al.} 2021,
  \mnras, 508, 3877

\bibitem[{{Graham} {et~al.}(2019){Graham}, {Kulkarni}, {Bellm}, {Adams},
  {Barbarino}, {Blagorodnova}, {Bodewits}, {Bolin}, {Brady}, {Cenko}, {Chang},
  {Coughlin}, {De}, {Eadie}, {Farnham}, {Feindt}, {Franckowiak}, {Fremling},
  {Gezari}, {Ghosh}, {Goldstein}, {Golkhou}, {Goobar}, {Ho}, {Huppenkothen},
  {Ivezi{\'c}}, {Jones}, {Juric}, {Kaplan}, {Kasliwal}, {Kelley}, {Kupfer},
  {Lee}, {Lin}, {Lunnan}, {Mahabal}, {Miller}, {Ngeow}, {Nugent}, {Ofek},
  {Prince}, {Rauch}, {van Roestel}, {Schulze}, {Singer}, {Sollerman}, {Taddia},
  {Yan}, {Ye}, {Yu}, {Barlow}, {Bauer}, {Beck}, {Belicki}, {Biswas}, {Brinnel},
  {Brooke}, {Bue}, {Bulla}, {Burruss}, {Connolly}, {Cromer}, {Cunningham},
  {Dekany}, {Delacroix}, {Desai}, {Duev}, {Feeney}, {Flynn}, {Frederick},
  {Gal-Yam}, {Giomi}, {Groom}, {Hacopians}, {Hale}, {Helou}, {Henning},
  {Hover}, {Hillenbrand}, {Howell}, {Hung}, {Imel}, {Ip}, {Jackson}, {Kaspi},
  {Kaye}, {Kowalski}, {Kramer}, {Kuhn}, {Landry}, {Laher}, {Mao}, {Masci},
  {Monkewitz}, {Murphy}, {Nordin}, {Patterson}, {Penprase}, {Porter},
  {Rebbapragada}, {Reiley}, {Riddle}, {Rigault}, {Rodriguez}, {Rusholme}, {van
  Santen}, {Shupe}, {Smith}, {Soumagnac}, {Stein}, {Surace}, {Szkody}, {Terek},
  {Van Sistine}, {van Velzen}, {Vestrand}, {Walters}, {Ward}, {Zhang}, \&
  {Zolkower}}]{2019PASP..131g8001G}
{Graham}, M.~J., {Kulkarni}, S.~R., {Bellm}, E.~C., {et~al.} 2019, \pasp, 131,
  078001

\bibitem[{Green(2023)}]{green2023}
Green, M. 2023, A Catalogue of All Known AM CVn Binary Systems

\bibitem[{{Green} {et~al.}(2020){Green}, {Marsh}, {Carter}, {Steeghs},
  {Breedt}, {Dhillon}, {Littlefair}, {Parsons}, {Kerry}, {Gentile Fusillo},
  {Ashley}, {Bours}, {Cunningham}, {Dyer}, {G{\"a}nsicke}, {Izquierdo}, {Pala},
  {Pattama}, {Outmani}, {Sahman}, {Sukaum}, \& {Wild}}]{green2020}
{Green}, M.~J., {Marsh}, T.~R., {Carter}, P.~J., {et~al.} 2020, \mnras, 496,
  1243

\bibitem[{{Green} {et~al.}(2018){Green}, {Marsh}, {Steeghs}, {Kupfer},
  {Ashley}, {Bloemen}, {Breedt}, {Campbell}, {Chakpor}, {Copperwheat},
  {Dhillon}, {Hallinan}, {Hardy}, {Hermes}, {Kerry}, {Littlefair}, {Milburn},
  {Parsons}, {Prasert}, {van Roestel}, {Sahman}, \&
  {Singh}}]{2018MNRAS.476.1663G}
{Green}, M.~J., {Marsh}, T.~R., {Steeghs}, D.~T.~H., {et~al.} 2018, \mnras,
  476, 1663

\bibitem[{{Kato} \& {Kojiguchi}(2021)}]{2021PASJ...73.1375K}
{Kato}, T. \& {Kojiguchi}, N. 2021, \pasj, 73, 1375

\bibitem[{{Kov{\'a}cs} {et~al.}(2002){Kov{\'a}cs}, {Zucker}, \&
  {Mazeh}}]{kovacs2002}
{Kov{\'a}cs}, G., {Zucker}, S., \& {Mazeh}, T. 2002, \aap, 391, 369

\bibitem[{{Kupfer} {et~al.}(2023){Kupfer}, {Korol}, {Littenberg}, {Shah},
  {Savalle}, {Groot}, {Marsh}, {Le Jeune}, {Nelemans}, {Petiteau}, {Ramsay},
  {Steeghs}, \& {Babak}}]{2023arXiv230212719K}
{Kupfer}, T., {Korol}, V., {Littenberg}, T.~B., {et~al.} 2023, arXiv e-prints,
  arXiv:2302.12719

\bibitem[{{Levitan} {et~al.}(2015){Levitan}, {Groot}, {Prince}, {Kulkarni},
  {Laher}, {Ofek}, {Sesar}, \& {Surace}}]{2015MNRAS.446..391L}
{Levitan}, D., {Groot}, P.~J., {Prince}, T.~A., {et~al.} 2015, \mnras, 446, 391

\bibitem[{{Masci} {et~al.}(2023){Masci}, {Laher}, {Rusholme}, {Shupe},
  {Paladini}, {Groom}, {Wold}, {Miller}, \& {Drake}}]{2023arXiv230516279M}
{Masci}, F.~J., {Laher}, R.~R., {Rusholme}, B., {et~al.} 2023, arXiv e-prints,
  arXiv:2305.16279

\bibitem[{{Masci} {et~al.}(2019){Masci}, {Laher}, {Rusholme}, {Shupe}, {Groom},
  {Surace}, {Jackson}, {Monkewitz}, {Beck}, {Flynn}, {Terek}, {Landry},
  {Hacopians}, {Desai}, {Howell}, {Brooke}, {Imel}, {Wachter}, {Ye}, {Lin},
  {Cenko}, {Cunningham}, {Rebbapragada}, {Bue}, {Miller}, {Mahabal}, {Bellm},
  {Patterson}, {Juri{\'c}}, {Golkhou}, {Ofek}, {Walters}, {Graham}, {Kasliwal},
  {Dekany}, {Kupfer}, {Burdge}, {Cannella}, {Barlow}, {Van Sistine}, {Giomi},
  {Fremling}, {Blagorodnova}, {Levitan}, {Riddle}, {Smith}, {Helou}, {Prince},
  \& {Kulkarni}}]{2019PASP..131a8003M}
{Masci}, F.~J., {Laher}, R.~R., {Rusholme}, B., {et~al.} 2019, \pasp, 131,
  018003

\bibitem[{{Ramsay} {et~al.}(2018){Ramsay}, {Green}, {Marsh}, {Kupfer},
  {Breedt}, {Korol}, {Groot}, {Knigge}, {Nelemans}, {Steeghs}, {Woudt}, \&
  {Aungwerojwit}}]{2018A&A...620A.141R}
{Ramsay}, G., {Green}, M.~J., {Marsh}, T.~R., {et~al.} 2018, \aap, 620, A141

\bibitem[{{Rodriguez} {et~al.}(2023){Rodriguez}, {Galiullin}, {Gilfanov},
  {Kulkarni}, {Khamitov}, {Bikmaev}, {van Roestel}, {Yungelson}, {El-Badry},
  {Sunayev}, {Prince}, {Buntov}, {Caiazzo}, {Drake}, {Gorbachev}, {Graham},
  {Gumerov}, {Irtuganov}, {Laher}, {Masci}, {Medvedev}, {Purdum},
  {Sakhibullin}, {Sklyanov}, {Smith}, {Szkody}, \&
  {Vanderbosch}}]{2023ApJ...954...63R}
{Rodriguez}, A.~C., {Galiullin}, I., {Gilfanov}, M., {et~al.} 2023, \apj, 954,
  63

\bibitem[{{Solheim}(2010)}]{2010PASP..122.1133S}
{Solheim}, J.~E. 2010, \pasp, 122, 1133

\bibitem[{{Szkody} {et~al.}(2020){Szkody}, {Dicenzo}, {Ho}, {Hillenbrand}, {van
  Roestel}, {Ridder}, {DeJesus Lima}, {Graham}, {Bellm}, {Burdge}, {Kupfer},
  {Prince}, {Masci}, {Mr{\'o}z}, {Golkhou}, {Coughlin}, {Cunningham}, {Dekany},
  {Graham}, {Hale}, {Kaplan}, {Kasliwal}, {Miller}, {Neill}, {Patterson},
  {Riddle}, {Smith}, \& {Soumagnac}}]{2020AJ....159..198S}
{Szkody}, P., {Dicenzo}, B., {Ho}, A. Y.~Q., {et~al.} 2020, \aj, 159, 198

\bibitem[{{Szkody} {et~al.}(2021){Szkody}, {Olde Loohuis}, {Koplitz}, {van
  Roestel}, {Dicenzo}, {Ho}, {Hillenbrand}, {Bellm}, {Dekany}, {Drake}, {Duev},
  {Graham}, {Kasliwal}, {Mahabal}, {Masci}, {Neill}, {Riddle}, {Rusholme},
  {Sollerman}, \& {Walters}}]{2021AJ....162...94S}
{Szkody}, P., {Olde Loohuis}, C., {Koplitz}, B., {et~al.} 2021, \aj, 162, 94

\bibitem[{{Szkody} \& {Van Roestel}(2023)}]{2023AAS...24145406S}
{Szkody}, P. \& {Van Roestel}, J. 2023, in American Astronomical Society
  Meeting Abstracts, Vol.~55, American Astronomical Society Meeting Abstracts,
  454.06

\bibitem[{{Toloza} {et~al.}(2019){Toloza}, {Breedt}, {De Martino}, {Drake},
  {Gansicke}, {Green}, {Ederoclite}, {Jennifer}, {Juna}, {Knigge}, {Kupfer},
  {Long}, {Marsh}, {Pala}, {Parsons}, {Prince}, {Raddi}, {Rebassa-Mansergas},
  {Rodriguez-Gil}, {Scaringi}, {Schmidtobreick}, {Schreiber}, {Schwope},
  {Shen}, {Steeghs}, {Szkody}, {Tappert}, {Toonen}, {Townsley}, \&
  {Zorotovic}}]{2019BAAS...51c.168T}
{Toloza}, O., {Breedt}, E., {De Martino}, D., {et~al.} 2019, \baas, 51, 168

\bibitem[{{van der Walt} {et~al.}(2019){van der Walt}, {Crellin-Quick}, \&
  {Bloom}}]{vanderWalt2019}
{van der Walt}, S., {Crellin-Quick}, A., \& {Bloom}, J. 2019, The Journal of
  Open Source Software, 4, 1247

\bibitem[{{van Roestel} {et~al.}(2021){van Roestel}, {Creter}, {Kupfer},
  {Szkody}, {Fuller}, {Green}, {Rich}, {Sepikas}, {Burdge}, {Caiazzo},
  {Mr{\'o}z}, {Prince}, {Duev}, {Graham}, {Shupe}, {Laher}, {Mahabal}, \&
  {Masci}}]{2021AJ....162..113V}
{van Roestel}, J., {Creter}, L., {Kupfer}, T., {et~al.} 2021, \aj, 162, 113

\bibitem[{{van Roestel} {et~al.}(2022){van Roestel}, {Kupfer}, {Green}, {Wong},
  {Bildsten}, {Burdge}, {Prince}, {Marsh}, {Szkody}, {Fremling}, {Graham},
  {Dhillon}, {Littlefair}, {Bellm}, {Coughlin}, {Duev}, {Goldstein}, {Laher},
  {Rusholme}, {Riddle}, {Dekany}, \& {Kulkarni}}]{2022MNRAS.512.5440V}
{van Roestel}, J., {Kupfer}, T., {Green}, M.~J., {et~al.} 2022, \mnras, 512,
  5440

\end{thebibliography}

\end{document}